\documentclass[aps,a4paper,10pt,twocolumn]{revtex4}
\usepackage{amsfonts}
\usepackage{amssymb}
\usepackage{mathrsfs}
\usepackage[mathscr]{euscript}
\usepackage[dvips]{graphicx}
\usepackage{fancyhdr}
\usepackage[hypertex=true]{hyperref}

\def\overstrike#1#2{{\setbox0\hbox{$#2$}\hbox to \wd0{\hss
    $#1$\hss}\kern-\wd0\box0}}

\begin{document}

\title{Phase sensitivity of perturbative nonlinear interactions}
\author{P. Kinsler}
\affiliation{
  Department of Physics, Imperial College,
  Prince Consort Road,
  London SW7 2BW, 
  United Kingdom.}
\author{G.H.C. New}
\affiliation{
  Department of Physics, Imperial College,
  Prince Consort Road,
  London SW7 2BW, 
  United Kingdom.}
\author{J.C.A. Tyrrell}
\affiliation{
  Department of Physics, Imperial College,
  Prince Consort Road,
  London SW7 2BW, 
  United Kingdom.}

\begin{abstract}

Despite the current concentration on phase control in few-cycle pulses,
 it emerges that there exists a wide class of nonlinear optical interactions
 in which the carrier phase is essentially irrelevant, 
 even for the shortest pulse profiles.  
Most parametric processes and most perturbative processes 
 fall into this category, 
 although others such as above threshold ionization (ATI) do not.
In an envelope approach, 
 the carrier oscillations are not part of the problem
 because they are removed at the outset.  
When they are reinstated at the end of the calculation, 
 one is free to include arbitrary phase shifts -- 
 within certain constraints.  
In many cases the constraints are relatively weak, 
 and it follows that a single envelope solution can be used with an
 infinite range of choices for the carrier phase.  

\end{abstract}

\pacs{X}


\newcommand{\sech}{{\textrm{ sech}}}

\lhead{\includegraphics[height=5mm,angle=0]{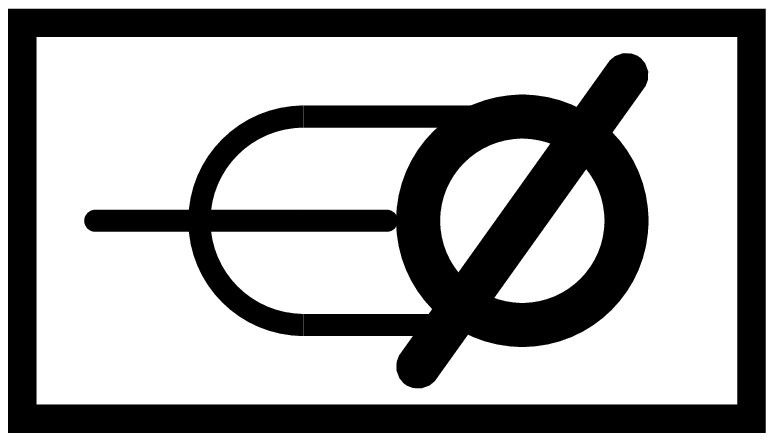}~~PHNLO}
\chead{~}
\rhead{
\href{mailto:Dr.Paul.Kinsler@physics.org}{Dr.Paul.Kinsler@physics.org}\\
\href{http://www.kinsler.org/physics/}{http://www.kinsler.org/physics/}
}

\date{\today}
\maketitle
\thispagestyle{fancy}


\chead{Phase sensitivity of PNL interactions}


\section{Introduction}\label{s-intro}

We have investigated the role that the carrier phase plays in the 
 propagation and interaction of few-cycle pulses affected by 
 perturbative nonlinear interactions, 
 using theoretical methods as well
 as envelope approaches and numerical solutions of Maxwell's equations.
{\em
The purpose of this paper is to examine in detail the
 process and consequences of how 
 perturbative nonlinear interactions become phase sensitive, 
 a subject usually passed over with only brief comment, 
 or in rather general terms.
}

Here focus our attention on $\chi^{(n)}$ nonlinear processes 
 and the interaction of multiple field components.  
Some of these,
 such as self phase modulation (SPM) are obviously phase insensitive.
Others, 
 are phase sensitive, 
 such as in the experiments of Jones et.al. \cite{Jones-DRSWHC-2000s} 
 who use interference with a super-continuum generated from 
 part of the input pulse, 
 or Morgner et.al. \cite{Morgner-EMSKFHI-2001prl} who use frequency doubling.  
Whether or not a $\chi^{(n)}$ process is carrier phase insensitive 
 depends both on the nonlinearity and the number of field components involved; 
 it is not guaranteed merely because it involves 
 optical pulses only a few cycles long.
The reason interactions involving many-cycle pulses are regarded 
 as phase insensitive
 has little to do with the role of the carrier phase, 
 but is instead due to the indistinguishability of the 
 many similar field oscillations in such pulses.  
Phase sensitive $\chi^{(2)}$ interactions involving degenerate 
 parametric soliton interactions have also been discussed 
 \cite{Assanto-1995ol,Drummond-KH-1999job}.  
Many non-perturbative processes sugh as above-threshold ionization (ATI) 
 (see e.g.  Apolonski et. al.\cite{Apolonski-PTSUHHK-2000prl}) are
 also phase sensitive, 
 however these are outside the intended scope of this paper.

We show 
 why  some perturbative nonlinear processes are phase insensitive, 
 but others are phase sensitive by reference to how is the  $\chi^{(n)}$ 
 processes are modelled, 
 and whether the various frequency and 
 nonlinear polarization terms are or are not significant.
By reference to the procedure used when 
 an envelope approach is adopted, 
 we systematize the procedure for generating models, 
 and can therefore clarify how phase (in)sensitivity occurs.
The key feature of an envelope approach \cite{Gabor-1946jiee} is that
 the carrier oscillations are removed at the outset and the presence of the
 carrier is not part of the solution.  
In the many cases where the carrier phase is irrelevant, 
 arbitrary phase shifts can be included 
 when the carriers are reinstated at the end of the calculation.  
It follows that a single envelope solution
 can include an infinite range of possibilities for the carrier phase.  
It is important to stress that this argument remains valid even if 
 the pulses are so short that attempting to find a solution based on 
 an envelope approach is impractical.  

In order to illustrate the results discussed in this paper, 
 we compare two types of simulation.  
Firstly, 
 we do envelope-based simulations of pulse propagation, 
 which naturally agree with the phase insensitivity predictions of 
 the envelope-based theory. 
Since we wish to  emphasize that there is nothing intrinsically many-cycle 
 or slowly-varying about these predictions, 
we propagate our pulse envelopes using the
 generalized few-cycle envelope approximation (GFEA) theory 
 \cite{Kinsler-N-2003pra,Kinsler-N-2004pra,Kinsler-2002arXiv-fcpp}.
This allows us (for example) to accurately model the self steepening 
 of the pulse envelope in self phase modulation (SPM).  
Secondly, 
 we do pseudospectral space-domain (PSSD) simulations \cite{Tyrrell-KN-2005jmo} 
 that solve Maxwell's equations without extra approximation. 
By varying the position of the carrier-like oscillations of 
 the PSSD initial conditions, 
 we are able construct a ``pseudo-envelope'' at each frequency of interest --  
 the maximum excursions of a set of PSSD simulations of varying phases.  
 For a set of results that match a phase insensitive envelope calculation, 
 the pseudo-envelope should mimic the true envelope.

\section{Background}

When using an envelope theory 
 (e.g. SVEA or GFEA) to solve a pulse propagation problem, 
 we find the carrier exponential (along with its phase $\phi$) 
 often vanishes from the description, 
 leaving only the chosen wavevector $k$ and frequency $\omega$
 as propagation parameters.   
The  question we now ask is: 
 {\em Under what conditions can we apply a carrier with a different $\phi$
 to the final envelope, 
 and still reconstruct a physically correct answer?}  
If we can change the underlying carrier phase, 
 our propagation is {\em phase insensitive}, 
 and for each envelope solution we can use an infinite set of
 phase choices. 
We will have solved for an infinite set of input electric field
 pulses with a single envelope propagation.

To examine this, 
 we need to consider the definition of a pulse
 envelope and carefully consider the role of the carrier phase.
An optical pulse can be written in the form
~
\begin{eqnarray}
  E(t, z) 
&=& 
  A(t,z) e^{\imath \left(k z - \omega t + \phi \right)} 
 +
  A^*(t,z) e^{-\imath \left(k z - \omega t + \phi \right)} 
,
\end{eqnarray}
 with \cite{Gabor-1946jiee} a carrier $e^{\imath
 \left( kz -\omega t + \phi \right)}$ 
 and an envelope $A(t,z)$.
A shift $\theta$ in the phase of this pulse 
 could be effected by either 
 adjusting the carrier (adding the shift $\theta$ to its existing 
 reference phase $\phi$); 
 or by multiplying the envelope $A(t,z)$ by $e^{\imath \theta}$.
We always choose the latter option, 
 so when the phase of the pulse evolves during its propagation, 
 it is the phase of the envelope $A$
 that changes whilst the carrier remains fixed. 
If, 
 for example, 
 the centre of the pulse (where $k z=\omega t$) evolves from 
 a cosine-like form at $z$ to a sine-like form at $z'$, 
 the sole effect is on the envelope, i.e. 
 $A(t,z) \longrightarrow A(t,z') = \imath A(t,z)$.

There is nothing intrinsically many (or few) cycle about 
 our conclusions here. 
If we can solve for the envelope propagation {\em in principle}, 
 then our predictions for phase sensitivity will still hold.  
Of course some of the contributions neglected by the SVEA and GFEA theories 
 do depend on the carrier phase,  
 but in general, 
 for most materials, 
 GFEA propagation is remarkably robust even for pulses containing only 
 a few carrier cycles.

The $\chi^{(n)}$ processes we consider here model 
 a wide variety of nonlinear optical behaviour.  
For these, 
 nonlinear polarization 
 affecting the pulse propagation of multiple field components 
 is a simple product of the sum of those components
~
\begin{eqnarray}
  P(t) 
&=&
  \chi^{(n)}
  \left[ \sum E_j(t) \right]^n
\\
&=& 
  \chi^{(n)}
  \left[ 
    \sum 
      \left( 
        A_j(t)  e^{ \imath \left( k_j z-\omega_j t +\phi_j \right)}
       + \textrm{c.c.}
      \right) 
  \right]^n
,
\end{eqnarray} 
 where $k_j, \omega_j$, and $\phi_j$ are chosen to have convenient values,
 depending on the details of the system being described.  
When expanded, 
 this will be a sum of many terms, 
 each of which comprises a product of envelope contributions and 
 carrier contributions.   
The normal procedure is to take each term in sequence, 
 combining all its carrier contributions into a single exponential.  
We then match the time-like behaviour of this to one of the
 individual carrier exponentials \cite{footnote-timematch} -- 
 so the envelope associated with that carrier propagates according to 
 the nonlinear contribution that best matches its frequency of oscillation. 
This frequency match will often be exact because of the way we have chosen 
 our set of carrier frequencies $\omega_j$, 
 but usually wavevector or carrier phase mismatches will remain.  
In most common situations,
 this carrier phase mismatch can be set equal to the phase of 
 the frequency matched carrier, 
 whilst still leaving some freedom in the overall choice of
 carrier phases.  

Note that the term ``phase mismatch'' is commonly used 
 in nonlinear optics to denote the phase evolution resulting from 
 a {\em wavevector} mismatch, 
 and does not refer to a carrier phase mismatch.  
To avoid the potential confusion generated by this widespread terminology, 
 in this paper we will only refer to either carrier phase mismatches 
 or wavevector mismatches, as appropriate.

As we will see, some nonlinear effects impose no relationship between the
 phases $\phi_j$ of the field carriers; 
 some allow a degree of freedom; 
 and others leave no freedom at all.  
Any  freedom to choose the $\phi_j$ in a particular case means that 
 that case is {\em phase insensitive}.  
One important restriction on the ability of $\chi^{(n)}$ interactions 
 to be phase insensitive is that they need to be (effectively) instantaneous, 
 so that the carrier oscillations at $\omega_j t$ 
 do not get folded into the calculation. 
For most nonlinear materials this is not a very significant restriction, 
 as the nonlinearities are very fast compared to 
 the optical frequencies of interest, 
 but in the case of (e.g.) slower semiconductor nonlinear materials
 this may remain relevant.

How these $\chi^{(n)}$ interactions stay phase insensitive (or not)
will become clearer after looking at the specific examples which
follow.


\section{Simple Cases}\label{S-simple}


A simple and commonly studied case of a nonlinear interaction is 
 self-phase modulation (SPM). 
The underlying nonlinearity that gives rise to SPM
 also gives rise to other terms, 
 such as third harmonic generation (THG) terms ($\sim A_i^3$) 
 and other cross terms (e.g. $\sim A_i A_j^{*2}$),
Although SPM is phase insensitive, 
 these other terms are not necessarily phase insensitive.  
In many cases, however, 
 they will be small and/or not wavevector matched -- 
 and if we are interested specifically in SPM, 
 then our setup will be designed to {\em ensure} they are negligible.

\begin{figure}
\includegraphics[width=40mm,angle=-90]{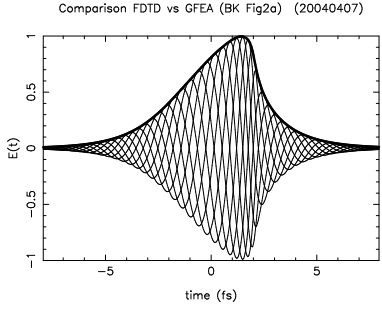}
\caption{
\label{F-SPM}
SPM: 
We match a GFEA envelope to a set of  PSSD 
solutions to Maxwell's equations, and achieve excellent agreement. 
It is also possible 
to use the envelope to generate field profiles that match the individual 
PSSD solutions. \\
{\bf Heavy line:} GFEA envelope solution.\\
{\bf Light lines:} PSSD fields, phase 
intervals of $\pi/4$. 
}
\end{figure}

The polarization contribution that affects the field envelope $A$ during 
 propagation under SPM is the ``polarization envelope'' $B_{SPM}$, 
 which has the same carrier as the field envelope. 
It is
~
\begin{eqnarray}
B_{SPM}
&=&
  \chi^{(3)} 
        A
      . A^*
      . A
,
\end{eqnarray} 
 which has no dependence on the carrier phase. 
We demonstrate this in fig \ref{F-SPM}, 
 where a set of PSSD solutions 
 of Maxwell's equations \cite{Tyrrell-KN-2005jmo} for SPM with
 a range of phases match up exactly with the phase insensitive GFEA envelope.  
The use of a GFEA\cite{Kinsler-N-2003pra} 
 rather than an SVEA pulse propagation equation for the envelope
 allows us to accurately reproduce the strong self-steepening that occurs for
 few-cycle pulses.  
It also emphasizes that this phase insensitivity still
 persists in the few-cycle regime.  

Another simple case is $n$-th order harmonic generation.  
This needs to be modeled using two field components, 
 with frequencies $\omega_n=n \omega_1$.
The underlying nonlinearity gives us a total polarization 
 $P(t) = \chi^{(n)} E(t)^n$, 
 from which the two important contribution $B_1$, $B_n$ affect 
 the propagation of the field envelopes $A_1, A_n$ respectively. 
As for the SPM case, 
 other polarization contributions might in some general case be significant, 
 but a harmonic generation experiment would be designed 
 to ensure they are negligible. 
The polarization envelopes $B_1$ and $B_n$  that match to the fundamental 
 $A_1$ and $n$-th harmonic  $A_n$ are 
~
\begin{eqnarray}
  B_1 
&=& 
  \chi^{(n)} 
  A_n {A_1^*}^{n-1} 
  \exp \left\{ \imath 
      \left[ 
        \Delta_n z
        + \phi_n - n \phi_1 
      \right]
    \right\}
,
\\
  B_n 
&=& 
  \chi^{(n)} 
  A_1^n 
  \exp \left\{ \imath 
      \left[ 
        - \Delta_n z
        +n \phi_1 - \phi_n 
      \right]
    \right\}
.
\end{eqnarray} 

Here $\Delta_n = k_n - n k_1$ is the wavevector mismatch.
These have no dependence on $\phi_k$'s 
 if $\delta = \phi_n - n \phi_1$ is fixed. 
This means there is a phase sensitivity, 
 but it only fixes the {\em relative} phase between the field components.  
This leaves one free choice of carrier phase,  
 and so the interaction can still be described 
 as phase insensitive.

\begin{figure}
\includegraphics[width=40mm,angle=-90]{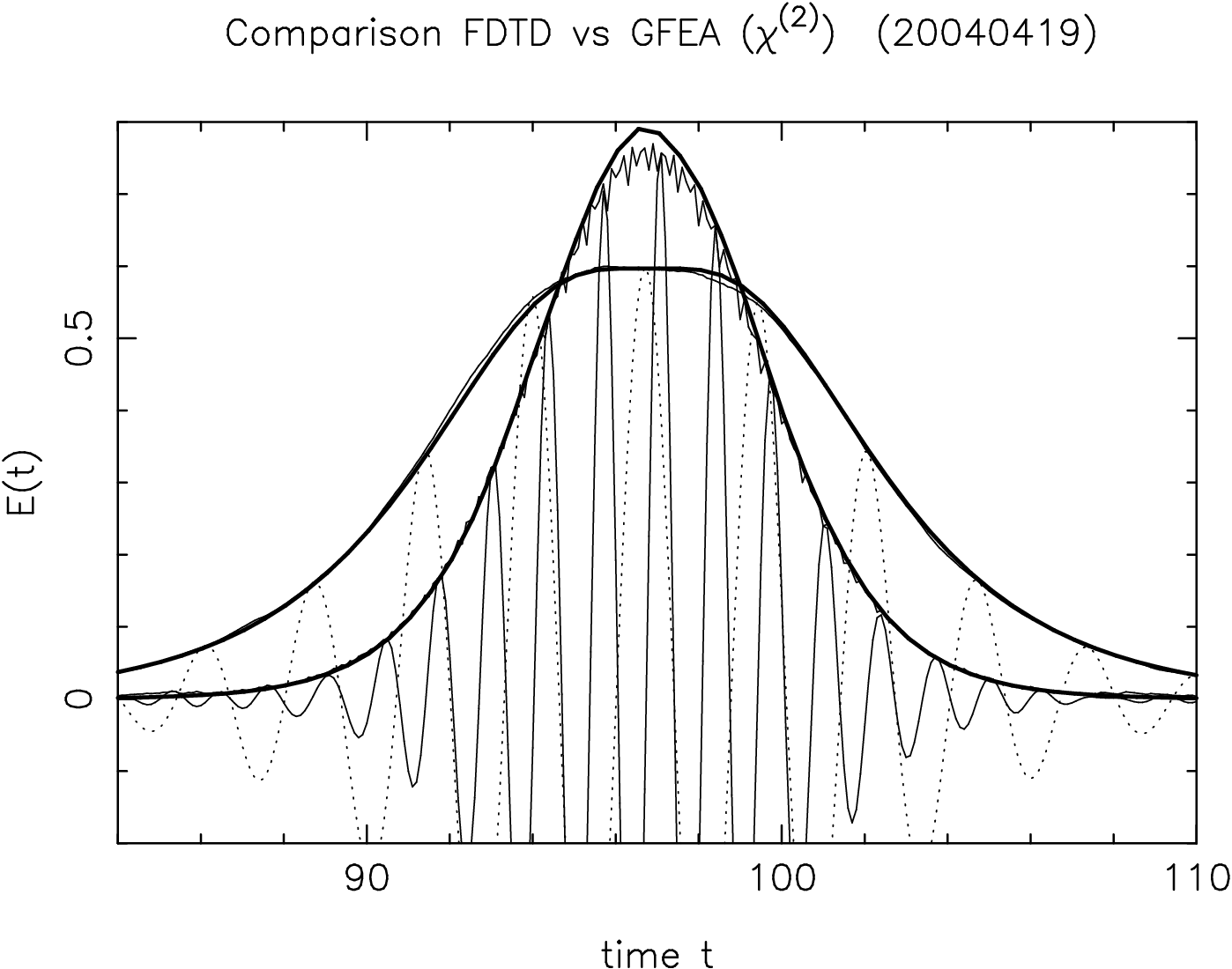}
\caption{
\label{F-CHI2-wband}
We match GFEA envelopes to a set of PSSD
solutions to Maxwell's equations with excellent agreement. \\
{\bf Heavy lines:} GFEA envelopes $A_1$ and $A_2$;\\
{\bf Light jagged lines:} PSSD pseudo-envelopes ;\\
{\bf Oscillating lines:} PSSD second harmonic field $E_2$
for $\phi_1=0$ (light), and the fundamental field $E_1$ (dotted).  \\
The second harmonic envelope curve does not exactly match the
pseudo-envelope because of both phase sampling and spectral filtering.  The
fundamental envelope and pseudo-envelopes barely be distinguished on the scale
of this graph.  
}
\end{figure}

To demonstrate the phase properties for harmonic generation we present some
 simulation results for the special case of second harmonic generation, 
 using two field components with frequencies $\omega_2=2 \omega_1$; 
 hence $\Delta_2 = k_2 - 2 k_1$ and 
 the carrier phase condition is $\phi_2 = 2 \phi_1$.  
In fig. \ref{F-CHI2-wband} we show results with 
 the initial condition $A_2=0$, 
 automatically satisfying the carrier phase condition, 
 but still allowing a free choice of the carrier phase $\phi_1$.  
Consequently, 
 it shows how a set of PSSD solutions (with range of initial $\phi_1$) 
 closely match the phase insensitive GFEA envelopes.
So, by ensuring that $\phi_2 = 2 \phi_1$, a single envelope calculation
 can give us answers for any choice $\phi_1$ value.

In these simulations we needed to separate 
 fundamental and second harmonic fields (and hence their pseudo-envelopes) 
 from the PSSD solutions for the total field.  
To do this we transformed into the spectral domain, 
 separated out the fundamental and second harmonic spectral peaks, 
 and transformed each back independently.  
This process involved some judgment; 
 and also filtered out the low frequency optical-rectification effects, 
 which were not included in the GFEA simulations in any case.  
This means that the resulting pseudo-envelopes seen on the 
 graphs are not perfect matches to the envelopes, 
 since, for example, 
 the field spectra contributions in the region between the carriers will not
 shared out in exactly the same way.

\begin{figure}
\includegraphics[width=60mm,angle=-90]{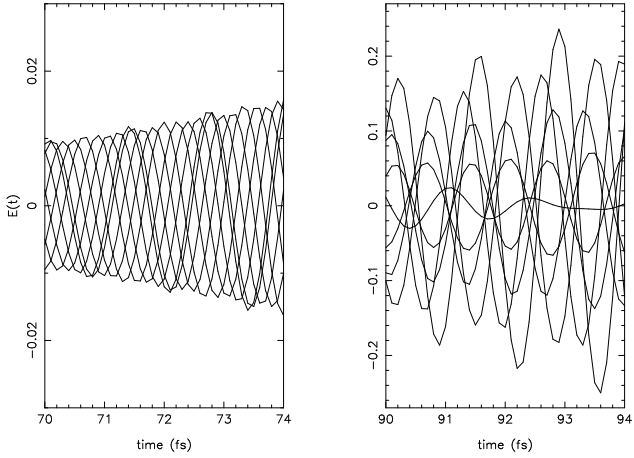}
\caption{
\label{F-CHI2-phi-E2multi2}
Two slices of the PSSD second harmonic field $E_2$ for a 
selection of $\phi_2$ values with $2 \phi_1 \neq \phi_2$. In the wings of 
the pulse (left hand, 70--74fs) 
we can see approximately envelope-like behavior because 
the phase-sensitive polarization terms are small; but as we move towards
the pulse 
peaks (right hand, 90--94fs), clearly no
reasonable envelope function could match all the data.\\
NB: peak $E_1(0)=1.00$ and peak $E_2(0)=0.20$.
}
\end{figure}

To contrast with the above results, 
 we now demonstrate  how {\em phase sensitivity} can manifest itself 
 in simulations that violate the $2 \phi_1 = \phi_2$ condition.  
We include a finite second harmonic pulse of varying phase $\phi_2$ in 
 the initial conditions, but fix $\phi_1=0$.  
On fig. \ref{F-CHI2-phi-E2multi} we show the case 
 where $\phi_2=0$, 
 which is the one phase that satisfies $2 \phi_1 = \phi_2$, 
 and as a result we see the GFEA envelope and PSSD fields in agreement.  
In contrast, on fig. \ref{F-CHI2-phi-E2multi2} 
 we show the second harmonic PSSD fields for a range of $\phi_2$, 
 and we see that no sensible envelope function could 
 reproduce the set of curves  -- 
 as expected since $2 \phi_1 \neq \phi_2$.

In a set of PSSD simulations where both $\phi_1$ and $\phi_2$ were varied, 
 a given GFEA simulation would reproduce the PSSD data along 
 a line where $2\phi_1=\phi_2$.  
Even if a particular case {\em is} phase insensitive, 
 it is not guaranteed to be of the type 
 we might find useful.

\begin{figure}
\includegraphics[width=40mm,angle=-90]{chi2-GFEA-E1E2}\\
\caption{
\label{F-CHI2-phi-E2multi}
We match the fundamental and second harmonic GFEA envelopes to the
corresponding PSSD
solution to Maxwell's equations, and see excellent agreement
($2 \phi_1=\phi_2=0$). \\
{\bf Heavy lines:} GFEA envelopes $A_1$ and $A_2$;\\
{\bf Light lines:} PSSD field split into $E_1$ and $E_2$.\\
NB: peak $E_1(0)=1.00$ and peak $E_2(0)=0.20$.
}
\end{figure}

\section{Complete Example: Three-field $\chi^{(2)}$ interaction}
\label{S-3field}

For a three-field $\chi^{(2)}$ interaction, 
 such as in an optical parametric amplifier (OPA),
 we get a number of polarization terms to consider.  
In contrast to the previous section, 
 which looked at common nonlinear polarization terms in isolation, 
 here we give a more comprehensive review of all the 
 nonlinear polarization terms in a particular case.
As we choose to use the interaction for an OPA or OPO system, 
 we follow the usual naming convention and 
 specify three field components (pump, signal, idler) with frequencies 
 $\omega_p > \omega_s \ge \omega_i$ such that $\omega_p=\omega_s+\omega_i$.
The polarization terms from $P(t) = E(t)^2$ that affect the 
 propagation of the field envelopes $A_p, A_s, A_i$ are -- 
~
\begin{eqnarray}
  P^+
&=& 
  A_p^2  e^{ \imath  2 \left( k_p z-\omega_p t +\phi_p \right)}
+
  A_p  e^{ \imath  \left( k_p z-\omega_p t +\phi_p \right)}
  A_s  e^{ \imath  \left( k_s z-\omega_s t +\phi_s \right)}
\nonumber 
\\
&&
+
  A_p e^{ \imath  \left( k_p z-\omega_p t +\phi_p \right)} 
  A_i e^{ \imath  \left( k_i z-\omega_i t +\phi_i \right)}
\nonumber 
\\
&&
+
  A_p    e^{ \imath  \left( k_p z-\omega_p t +\phi_p \right)}
  A_s^*  e^{-\imath  \left( k_s z-\omega_s t +\phi_s \right)}
\nonumber 
\\
&&
+
  A_p    e^{ \imath  \left( k_p z-\omega_p t +\phi_p \right)}
  A_i^*  e^{-\imath  \left( k_i z-\omega_i t +\phi_i \right)}
+
  A_s^2  e^{ \imath  2 \left( k_s z-\omega_s t +\phi_s \right)}
\nonumber 
\\
&&
+
  A_s  e^{ \imath  \left( k_s z-\omega_s t +\phi_s \right)}
  A_i  e^{ \imath  \left( k_i z-\omega_i t +\phi_i \right)}
\nonumber 
\\
&&
+
  A_s  e^{ \imath  \left( k_s z-\omega_s t +\phi_s \right)}
  A_i^*  e^{ -\imath  \left( k_i z-\omega_i t +\phi_i \right)}
\nonumber 
\\
&&
+
  A_i^2  e^{ \imath  2 \left( k_i z-\omega_i t +\phi_i \right)}
.
\end{eqnarray} 

The zero and negative frequency polarization components are not 
 included here -- 
 the negative frequency ones ($P^-$) apply to the evolution of 
 the conjugate parts of the envelopes $A_p^*, A_s^*, A_i^*$; 
 and the zero frequency ones ($P^0$) correspond to 
 the uninteresting (for us) steady state contributions.

In the following we will start by considering the three polarization 
 contributions that are commonly considered, 
 then step-by-step add the rest back in to the model, 
 describing the consequences for the phase sensitivity.

\subsection{Standard (resonant) case}

The {\em standard subset} of polarization contributions are 
 the ones exactly resonant with one of the three carriers.  
As usual, 
 $B_\alpha$ is a polarization envelope with the carrier 
 $\exp\left[ \imath  
 \left( k_\alpha z-\omega_\alpha t +\phi_\alpha \right) \right]$,
 so that with the wavevector mismatch $\Delta=k_p -k_s -k_i$, 
 we have
~
\begin{eqnarray}
  B_i 
= 
B_{i,R}
&=&
  A_p A_s^* 
  e^{ \imath  \Delta z 
     +\imath  \left( \phi_p -\phi_s -\phi_i\right) }
\\
  B_s
= 
B_{s,R}
&=&
  A_p A_i^* 
  e^{ \imath  \Delta z
     +\imath  \left( \phi_p -\phi_i -\phi_s\right) }
\\
  B_p
= 
B_{p,R}
&=&
  A_s A_i 
  e^{-\imath  \Delta z
     +\imath  \left( \phi_i +\phi_s -\phi_p\right) }
\end{eqnarray} 
~
Each gives us (the same) carrier phase constraint (A): 
 $\phi_p=\phi_s+\phi_i$, 
 and so any single envelope simulation can be turned into 
 a multiplicity of solutions for the fields, 
 using the two unconstrained carrier phases to determine the third.   
Thus, for resonant-only polarization contributions, 
the interaction is doubly phase insensitive 
 (i.e. we have two free choices of carrier phase).
 
\subsection{Wideband case}

Of course in a wideband (or miraculously well wavevector matched) situation,
 other polarization terms might be relevant. 
Assuming that $\omega_s-\omega_i \sim \omega_i$, 
 and $2 \omega_i \sim \omega_s$, 
 we would associate the extra polarization contributions with the 
 propagation of the envelope with the nearest carrier frequency, 
 so
~
\begin{eqnarray}
  B_i 
= 
  B_{i,R} 
 + 
  B_{i,W}
&=&
  B_{i,R} 
 + 
  A_s A_i^* 
  e^{ \imath  \left( k_s -2 k_i \right) z 
     +\imath  \left( \phi_s -2 \phi_i\right) }
~~~~
\label{eq-threefieldchi2-wideband-si}
\\
  B_s
= 
  B_{s,R} 
 + 
  B_{s,W}
&=&
  B_{s,R} 
 + 
  A_i^2 
  e^{ \imath  \left( 2 k_i -k_s \right) z 
     +\imath  \left( 2 \phi_i -\phi_s \right) }
~~~~
\label{eq-threefieldchi2-wideband-ii}
\end{eqnarray} 

Both of these add the extra phase constraint (B): $\phi_s=2\phi_i$. 
 Substituting this into constraint (A) above therefore gives us the new
 constraint $\phi_p=3\phi_i$; 
 which leaves us with just one unconstrained phase
 from which we can generate multiple field solutions from 
 a single envelope simulation.

For example,
 in some recent carrier-envelope phase stabilisation experiments
 \cite{Fang-K-2004ol,Baltuska-FK-2002prl},
 the authors advance an argument based (only) on the resonant terms
 to explain why they can generate an idler field insensitive to the 
 initial phase of the pump pulse (from which the signal is also 
 derived).
Since they have near resonant signal and idler fields, 
 the two \lq\lq wideband\rq\rq ~ terms above in 
 eqns. (\ref{eq-threefieldchi2-wideband-si}, 
        \ref{eq-threefieldchi2-wideband-ii})
 will produce components near the pump (since $2 \omega_i \sim \omega_p$) 
 and near the zero-frequency (since $\omega_s - \omega_i \sim 0$)
 respectively.
Consquently their model is very insensitive to these 
 contributions to the nonlinearity,
 especially when the pump field is strong.
However, 
 if they had chosen a scheme with $\omega_s-\omega_i \sim \omega_i$,
 then the $\omega_s-\omega_i$ nonlinear component would 
 disrupt their phase independent idler generation -- 
 although fortunately the term would be weak,  
 since it is dependent on the idler strength.
More serious problems would have ensued if the scheme required 
 generation of a signal pulse from pump-idler interaction, 
 since then a strong $\omega_i+\omega_i$ term would disrupt the 
 hoped-for  generation of a phase-independent signal pulse.

\subsection{Extreme case}

Finally, in some extreme cases we might need (or want) to include
 the remaining polarization terms.  
Because of their high frequency,
 they are best assigned to drive the $\omega_p$ field component
~
\begin{eqnarray}
  B_p
&=& 
  B_{p,R} 
 + 
  B_{p,X}
\\
  B_{p,X} 
&=&
  A_p
  A_i  
  e^{ \imath  
      \left[ \left(k_p+k_i\right) z
             -\left(\omega_p+\omega_i\right) t 
             +\phi_p+\phi_i 
      \right]
    }
\nonumber 
\\
&&
 +
  A_p
  A_s
  e^{ \imath  
      \left[ \left(k_p+k_s\right) z
             -\left(\omega_p+\omega_s\right) t 
             +\phi_p+\phi_s 
      \right]
    }
\nonumber 
\\
&&
 +
  A_s^2  e^{ \imath  2 \left( k_s z-\omega_s t +\phi_s \right)}
 +
  A_p^2  e^{ \imath  2 \left( k_p z-\omega_p t +\phi_p \right)}
.
~~~~
\end{eqnarray} 
~
These would add constraints $\phi_i=0$, $\phi_s=0$, $\phi_p=2\phi_s$, 
 and $\phi_p=0$ respectively, 
 any of which would leave no adjustable carrier phases.  
This means that a single envelope simulation corresponds to only 
 one single field result.  
Trying to avoid this by adding extra carriers 
 (e.g. by inventing a $\omega_q=2\omega_s$) 
 is generally counter productive, 
 as these will generate extra polarization terms, 
 each of which will arrive with a new 
 collection of constraints.

\section{High harmonic generation}\label{S-ati}

High harmonic generation is the process whereby an intense laser
 field is use to ionize an atom, 
 with the electron recollsion with
 the atom generating a wide range of high harmonics off the 
 laser pulse.  
It is clearly a  phase sensitive process, 
 because for a short (few cycle) laser pulse it is possible that only one of 
 the carrier-like field oscillations is sufficiently near the peak of 
 the pulse to reach the ionization threshold.  
Changing the relative position of the carrier w.r.t. 
 the pulse peak could then shift the timing of the 
 above-threshold part of the electric field profile, 
 and even possibly shift it from a 
 positive value electric field to a negative value.  
This will either alter the timing or direction of the ejected electron, 
 and hence its recollision and the generated HHG signal.
The typical description of HHG focuses on the electron trajectory,
 following a tunnel-ionization event (caused by the distortion
 of the atomic potential of the atom or molecule.  
Such models bear no relationship to the perturbative $\chi^{(n)}$ models 
 already discussed here,
 and so cannot be interpreted in the way we found useful above.

However, 
 since HHG is a multi photon process, 
 it is instructive to imagine a  high-order nonlinear
 ``perturbative HHG'' process which is compatible with our approach. 
We emphasize that here we are not trying to make an accurate model of 
 the HHG process, 
 but to reproduce some of its features in a way that enables us to 
 relate its phase sensitivity to that of 
 standard $\chi^{(n)}$ nonlinear processes.

In our simple model, 
 we consider just one high-order multi photon polarization term,
 and show how the phase constraints it generates remove any 
 possibility of phase freedom.
Considering an odd high-order process, 
 we specify $N=2m+1$ (i.e. $N$ odd), 
 and write down the polarization terms resulting from 
 a single envelope-carrier combination 
 at the fundamental frequency $\omega$ --
~
\begin{eqnarray}
P_{ATI}
  &=& 
  - \gamma
       E(t)^N
\\
&=&
  - \gamma
  \left[
    \sum_{n=0}^{N}
    C^{N}_n A^{*(N-n)} A^n 
        e^{\imath \left( 2n - N \right) 
                  \left( kz - \omega t + \phi \right)
          }
  \right] 
,
\end{eqnarray} 
 where $C^{N}_n$ are the binomial coefficients.
Thus the $n$-th polarization term drives the electron at
 a frequency of $\left(2n - N\right) \omega$  --
~
\begin{eqnarray}
  P_{2n - N}
&=&
  - \gamma ~
     C^{N}_n 
        ~ A^{*(N-n)} 
        ~ A^n ~
        e^{\imath \left( 2n - N  \right) 
                  \left( kz - \omega t + \phi \right)
          }
.
~~~~
\label{eqn-ati-polarizationenvelopes}
\end{eqnarray} 

If we compare the effect of the $n$-th term on the electron to that 
 of the $n'$-th term, 
 we see that they beat with an exponential dependence like 
~
\begin{eqnarray}
        e^{\imath \left( 2n - 2n'  \right) 
                  \left( kz - \omega t + \phi \right)
          }
.
~~~~
\label{eqn-ati-polarizationbeat}
\end{eqnarray} 

Even this first comparison shows a clear phase sensitivity; before we 
 have considered all $N$ of the terms, and before we consider the 
 likelihood of contributions from interactions of a different order (e.g. $N'$).

In the cases discussed in previous sections, 
 we could regard some terms as 
 negligible because they were small in magnitude or poorly wave-vector matched.
Here all terms contain $N$ powers of the envelope amplitude, 
 thus none can be dismissed as negligible in size.  
All, however, have different frequencies.  
Our comparisons above shows us that as long as 
 the electron can respond significantly to even just two terms,
 it's response will be phase sensitive.

\section{Conclusions}

We have shown that  nonlinear interactions can run the full range from 
 completely carrier phase insensitive through to completely constrained, 
 as demonstrated for   
 both the simple cases in section \ref{S-simple} and the detailed examination
 of a three-field $\chi^{(2)}$ interaction in section \ref{S-3field}.  
The principles and procedure outlined in section \ref{S-3field} 
 can easily be applied to higher order $\chi^{(n)}$ nonlinearities with 
 differing numbers of field components -- 
 all that is necessary is a careful enumeration of the field components and
 significant polarization terms, 
 followed by the process of matching up the polarization terms to 
 the evolution of the field components. 
Lastly, 
 in section \ref{S-ati} we saw how any high-order $\chi^{(n)}$ process will, 
 without carefully designed conditions, 
 be be phase sensitive purely because of the large number of constraints 
 generated by the polarization expansion.

Although strictly speaking all nonlinear
 $\chi^{(n)}$ interactions allow no residual carrier phase insensitivity, 
 such complete descriptions are rarely necessary for accurate physical models.  This is because in many realistic cases we can safely ignore many of 
 the nonlinear polarization contributions and so recover some phase freedom, 
 as  demonstrated in this paper.  
This is because the neglected terms will be far off resonance
 and/or not wavevector matched enough to accumulate and so will play a
 negligible part in the propagation.  

The phase (in)sensitivity of the particular models pursued 
 in this paper relate to experiments involving the phase-control of 
 few cycle pulses 
 (e.g.
 \cite{Apolonski-PTSUHHK-2000prl,Jones-DRSWHC-2000s,Morgner-EMSKFHI-2001prl}) 
 in the following way: 
 a scheme will be phase sensitive, 
 and hence useful to the experimentalists {\em if}, when analyzed, 
 there is no freedom to alter the carrier phase under 
 some predicted final-state pulse envelope.  
This contrasts with the case for processes like SPM, 
 where a single envelope propagation code can predict a final-state 
 envelope where {\em any} carrier phase can be used to generate a 
 valid final-state electric field profile.

The phase insensitivity we have discussed can manifest itself
 over the complete range from many- to few-cycle pulses -- 
 as long as {\em in principle} an envelope solution to the propagation 
 is possible -- 
 and {\em in principle}, 
 we might even retain all of the terms approximated away in 
 the GFEA pulse propagation theory. 
Our conclusions are supported by successful comparisons between 
 envelope predictions of carrier phase insensitivity and 
 sensitivity with PSSD solutions of Maxwell's equations.


\end{document}